\documentclass[aps, prb, 10pt, twocolumn, superscriptaddress,floatfix]{revtex4-1}

\usepackage{graphicx}
\usepackage{amsmath,amssymb,amsfonts}
\usepackage[percent]{overpic}
\usepackage{braket}
\usepackage{bm}
\usepackage[breaklinks=true,colorlinks,citecolor=blue,linkcolor=blue,urlcolor=blue]{hyperref}
\usepackage[usenames, dvipsnames]{color}
\begin{document}
\title{Many-body localized quantum batteries}

\author{Davide Rossini}
\email{davide.rossini@unipi.it}
\affiliation{Dipartimento di Fisica dell'Universit\`a di Pisa and INFN, Largo Pontecorvo 3, I-56127 Pisa,~Italy}

\author{Gian Marcello Andolina}
\affiliation{NEST, Scuola Normale Superiore, I-56126 Pisa,~Italy}
\affiliation{Istituto Italiano di Tecnologia, Graphene Labs, Via Morego 30, I-16163 Genova,~Italy}

\author{Marco Polini}
\affiliation{Istituto Italiano di Tecnologia, Graphene Labs, Via Morego 30, I-16163 Genova,~Italy}
\begin{abstract}
The collective and quantum behavior of  many-body systems may be harnessed to achieve fast charging of energy storage devices, which have been recently dubbed ``quantum batteries". In this Article, we present an extensive numerical analysis of energy flow in a quantum battery described by a disordered quantum Ising chain Hamiltonian, whose equilibrium phase diagram presents many-body localized (MBL), Anderson localized (AL), and ergodic phases. We demonstrate that i) the low amount of entanglement of the MBL phase guarantees much better work extraction capabilities, measured by the ergotropy, than the ergodic phase and ii) interactions suppress temporal energy fluctuations in comparison with those of the non-interacting AL phase. Finally, we show that the statistical distribution of values of the optimal charging time is a clear-cut diagnostic tool of the MBL phase.
\end{abstract}

\maketitle

\section{Introduction}

Recently, there has been a great deal of interest in studying charging times and work extraction in ``quantum batteries" (QBs)~\cite{Alicki13,Hovhannisyan13,Binder15,Campaioli17,Le17,Ferraro17,Andolina18,Andolina19,Farina18,Julia-Farre18,Zhang18,Andolina18c,zhang_arxiv_2018,Campaioli18,Kaneko19}. A QB is composed by $N$ identical quantum cells where energy is stored and from which work can be extracted (at least in principle). Prototypical examples of QBs that have been studied include arrays of $N$ qubits~\cite{Ferraro17,Andolina18,Andolina19,Farina18,Julia-Farre18,zhang_arxiv_2018} and XXZ Heisenberg spin chains with $N$ spins~\cite{Le17}.  The quantum cells are either coupled to an energy source, e.g. a photonic cavity~\cite{Ferraro17}, or they are charged via a non-equilibrium quantum quench~\cite{Polkovnikov11} after they have been prepared in a given state, physically representing the discharged battery. In the first instance, the coupling allows energy flow from the energy source into the battery. (In the case of Dicke QBs comprising $N$ qubits in a photonic cavity~\cite{Ferraro17}, the latter also mediates long-range interactions between the qubits.) In the second instance, the work needed to perform the quench charges the quantum cells during the non-equilibrium dynamics.

In a QB, one is interested in minimizing the charging time, exploiting the collective behavior of the ensemble of quantum cells and, possibly, genuine quantum features of the charging dynamics. At the same time, one needs to maximize the fraction of energy stored in the QB that can be extracted in order to perform thermodynamic work. If the energy stored in the battery is indeed locked by correlations between the $N$ quantum cells~\cite{Andolina19}, it cannot be extracted and, despite being maximal at the optimal charging time, is useless from the point of view of performing work.

In this Article, we consider a QB model, which, at equilibrium, displays a many-body localized (MBL) phase. MBL phases~\cite{Nandkishore15,Alet18,Abanin18} have been and still are at the center of intensive theoretical~\cite{Nandkishore15,Alet18,Abanin18} and experimental~\cite{Schreiber15,choi_science_2016,Bordia16,Smith16,Bordia17,Luschen17,Roushan17,wei_prl_2018,Xu18,Lukin19} investigations. Many-body localization~\cite{Basko06,Gornyi05} is a phenomenon occurring in an interacting disordered quantum many-particle system, in the regime where all single-particle eigenstates are localized by disorder. In condensed matter systems, it is known that electron-phonon collisions induce variable-range-hopping transport at low temperatures. Surprisingly, only recently it has been demonstrated~\cite{Basko06,Gornyi05} that electron-electron interactions are incapable of doing so and that, in the absence of phonons, the electron system is locked into an insulating state with zero dc electrical conductivity, up to a critical temperature $T_{\rm c}$. For $T> T_{\rm c}$, interactions produce a metallic state. Since for $T<T_{\rm c}$ electron-electron interactions alone are unable to establish thermal equilibrium, the study of many-body localization has greatly increased our knowledge of the approach to equilibrium in a closed quantum system, providing us with the only robust mechanism known so far to avoid thermalization~\cite{Nandkishore15,Alet18,Abanin18}. Another aspect of MBL states of interest to this work is the fact that these display a low amount of entanglement and obey~\cite{Nandkishore15,Alet18,Abanin18} an area-law rather than a volume-law. Indeed, the entanglement entropy of a subsystem in an MBL eigenstate scales proportionally to the volume of the boundary of the subsystem~\cite{eisert_rmp_2010}. This is in stark contrast to ergodic states, whose entanglement scales like the volume of the subsystem.

The reasons to investigate MBL states in the context of QBs are threefold. i) All quantum many-body battery models studied so far~\cite{Le17,Ferraro17,Andolina19,Julia-Farre18,Andolina18c,zhang_arxiv_2018} display ground-state quantum phase transitions~\cite{Sachdev}. The charging dynamics in these models is therefore qualitatively insensitive to these phase transitions since it involves highly excited states. On the other hand, many-body localization involves excited states with finite energy density and is thus expected to imprint qualitative features on the charging dynamics. ii) The low amount of entanglement carried by MBL states is expected~\cite{Andolina19} to lead to a {larger value of extractable work (ergotropy) from the QB}
in comparison with phases which obey a volume-law, like the ergodic one. iii) Finally, interactions in the MBL phase are expected to suppress temporal energy fluctuations with respect to other phases obeying an area-law, like the Anderson localized (AL) phase. In practice, this means much more stable batteries.

Below, we present the outcome of extensive numerical calculations on a specific QB model Hamiltonian, namely that of a disordered quantum Ising chain. These fully support the above expectations.
  Moreover, we show that the statistical distribution of optimal charging times of a QB provides a powerful diagnostic tool of MBL phases and, most importantly, a tool that does not rely on looking at the long-time dynamics of the many-particle system~\cite{Abanin18}.
  We emphasize that we have also simulated an alternative QB model, that of a disordered quantum Heisenberg spin chain. We have reached the same conclusions.
  This makes us reasonably confident on the general nature of our findings, in the context of models displaying MBL,
AL, and ergodic phases.

Our Article is organized as following. In Sect.~\ref{sect:protocol} we provide a general definition of a many-body QB and discuss a suitable quench protocol for its charging process, following the strategy outlined in Ref.~\onlinecite{Le17}.
  We also present the figures of merit we use in our analysis as quantifiers of the ``performance" of a many-body QB.
  The specific model Hamiltonian of our QB is then introduced in Sect.~\ref{sect:model}, together with a brief presentation of its phase diagram.
  Sect.~\ref{sect:results} contains the main results we obtained, and constitutes the core of the Article, while in Sect.~\ref{sect:discussion}
  we draw our main conclusions. 
  Five appendices deal with technical issues concerning the scaling of the energy stored in the QB with its size (App.~\ref{sec:scaling}),
  the influence of temporal fluctuations (App.~\ref{sect:temporal}) and that of the observation time window (App.~\ref{sect:dependence}).
  We also discuss the cases of a non-disordered spin-chain QB (App.~\ref{sect:ergotropy}) and of a QB modelled by a Heisenberg spin chain in a random magnetic field (App.~\ref{sect:heisenberg}).

\section {Many-body quantum batteries and figures of merit}
\label{sect:protocol}

\subsection{Charging protocol}
\label{subsec:QB}

We start by defining a quantum-quench-based protocol for the charging process of a QB.
We assume that our battery is made of $N$ identical quantum cells, which are governed by a free and local Hamiltonian~\cite{footnote1},
\begin{equation}
\mathcal{H}_0 = \sum_{j=1}^N {h}_j~.
\end{equation}
At time $t=0$, the system is prepared in its own ground state $|0\rangle$ (representing the discharged battery).
By suddenly switching on a suitable interaction Hamiltonian $\mathcal{H}_1$ for a finite amount of time $\tau$,
one aims to inject as much energy as possible into the quantum cells~\cite{Binder15,Campaioli17,Le17}.
The time interval $\tau$ is called the {\it charging time} of the protocol.

The full Hamiltonian of the model can be written as
\begin{equation}
  \label{eq:protocol}
  \mathcal{H}(t) = \mathcal{H}_{\rm 0}+\lambda(t)\mathcal{H}_1~,
\end{equation}
where $\lambda(t)$ is a classical parameter that represents the external
control exerted on the system, and which is assumed to be given by
a step function equal to $1$ for $t\in[0,\tau]$ and zero elsewhere. {(This fast switch can be implemented experimentally, e.g., by controlling an external magnetic field in a setup where the quantum cells are realized via superconducting qubits~\cite{Fink09}.)}
Accordingly, denoting by $|\psi(t) \rangle$ the evolved
state of the system at time $t$, its total energy
$E^{\rm tot}_N(t) = {\langle} \psi(t) |\mathcal{H}(t)| \psi(t) \rangle$
is constant at all times, with the exception of the switching points, $t=0$ and $t=\tau$.
The energy injected into the $N$ quantum cells can be thus expressed in terms of the mean local energy at the end of the protocol:
\begin{equation}
  E_N(\tau) = {\langle} \psi(\tau) |\mathcal{H}_0| \psi(\tau) \rangle~.\label{eq:stored energy}
\end{equation}
The optimal charging time $\bar{\tau}$ is defined by~\cite{footnote2}
\begin{equation}
E_N(\bar{\tau})=\max_{\tau>0} E_N({\tau})~.
\end{equation}

\subsection{Ergotropy}
\label{subsec:ergo}

An important quantity in the above protocol is the fraction of $E_N(\tau)$ that can be effectively extracted from $M\leq N$ quantum cells, without having access to the full system. Indeed, part of the energy injected into the battery is typically locked by correlations that are established between the $N$ quantum cells during the charging dynamics, and is thus useless for practical purposes~\cite{Andolina19}.

A proper measure of the fraction of $E_N(\tau)$ that can be utilized to perform thermodynamic work is provided by the so-called {\it ergotropy}~\cite{Allahverdyan04,Lenard1978,Pusz1978}
of the local state $\rho_{M}(\tau)$. Here, $\rho_{M}(\tau)$ is the reduced density matrix---evaluated at time $\tau$---of $M\leq N$ quantum cells, 
and governed by the Hamiltonian $\mathcal{H}^{(M)}_0=\sum_{j=1}^M {h}_j$.

We remind that, for a quantum system $\rm X$ characterized by a local Hamiltonian $\mathcal H$,
the ergotropy $\mathcal{E}(\rho,{\mathcal H})$ is a functional measuring the maximum amount of energy
that can be extracted from a density matrix $\rho$ of $\rm X$, using an arbitrary unitary operator.
A closed expression for this quantity can be obtained in terms of the difference
\begin{equation}
  \mathcal{E}(\rho,{\cal H})=E(\rho)-E(\tilde{\rho})
\end{equation}
between the mean energy $E(\rho) = {\rm tr}[\mathcal{H} \rho]$ of the state $\rho$ and that of
the passive counterpart $\tilde{\rho}$ of $\rho$~\cite{Lenard1978,Pusz1978,Allahverdyan04},
i.e.~$E(\tilde{\rho}) = {\rm tr}[\mathcal{H} \tilde{\rho}]$. 
The latter is defined as the density matrix of $\rm X$ which is diagonal on the eigenbasis
of ${\cal H}$ and whose eigenvalues correspond to a proper reordering
of those of $\rho$, i.e.,
\begin{equation}
  \tilde{\rho}=\sum_n r_n \ket{\epsilon_n}\bra{\epsilon_n}~,
\end{equation}
where 
\begin{eqnarray}
  \rho = & \sum_n r_n\ket{r_n}\bra{r_n}, \qquad & r_0\geq r_1 \geq r_2 \geq \cdots~, \\
  \mathcal{H} = & \sum_n \epsilon_n\ket{\epsilon_n}\bra{\epsilon_n}~, \qquad & \epsilon_0\leq \epsilon_1 \leq \epsilon_2 \leq \cdots~.
\end{eqnarray}
Therefore $E(\tilde{\rho})=\sum_{n} r_n \epsilon_n$.
Notice that, if $\epsilon_0=0$ and the state $\rho$ is pure, then $E(\tilde{\rho})=0$
and the ergotropy coincides with the mean energy of $\rho$, i.e.~$\mathcal{E}(\rho,{\cal H})=E(\rho)$.
On the contrary, if the state $\rho$ is mixed, the extractable work is in general smaller
than the mean energy, i.e.~$\mathcal{E}(\rho,{\cal H}) < E(\rho)$.

Since the global evolution is unitary, $\rho_{N}(t)= |\psi(t)\rangle \langle \psi(t)|$ remains pure at all times.
In contrast, the local state $\rho_{M}(\tau)$ will be in general mixed,
because of its entanglement with the remaining $N-M$ elements of the battery,
introducing a non-trivial ``gap'' between its ergotropy $\mathcal{E}_M(\tau) \equiv \mathcal{E} \big[ \rho_{M}(\tau),{\mathcal H}^{(M)}_{0} \big]$ and the energy ${E}_{M}(\tau)= {\rm tr} \big[ \mathcal{H}^{(M)}_0\rho_{M}(\tau) \big]$
stored in the subsystem at the end of the charging process, see Eq.~\eqref{eq:stored energy}.

\section{Model}
\label{sect:model}

We consider a spin-chain QB. In the absence of charging operations, the system is described by the following non-interacting Hamiltonian:
\begin{equation}
  \mathcal{H}_0 = h \sum_{j=1}^N \sigma^z_j~,
  \label{eq:HamStatic}
\end{equation}
where $h>0$ represents a transverse magnetic field (with units of energy), $\sigma^\alpha_j$ ($\alpha=x,y,z$) are standard spin-$1/2$ Pauli matrices, and $N$ is the number of spins (i.e.~our quantum cells) in the QB.
At time $t=0$ the system is initialized in the ground state
\begin{equation}
\ket{0} = \bigotimes_{j=1}^{N} \ket{\downarrow}_j~.
\end{equation}
In order to inject energy into the system, we perform a sudden quench
by switching on at time $t=0$ the Hamiltonian $\mathcal{H}_1$ in Eq.~\eqref{eq:protocol}. We take
\begin{equation}
  \mathcal{H}_1 = - \sum_{j=1}^N J_j \sigma^x_j \sigma^x_{j+1} + J_2 \sum_{j=1}^N \sigma^x_j \sigma^x_{j+2}~.
  \label{eq:HamB}
\end{equation}
The two terms entering $\mathcal{H}_1$ describe nearest-neighbor and next-to-nearest-neighbor spin-spin interactions. The couplings $J_j = J + \delta J_j$ are the sum of a fixed part $J$ and a randomly-varying contribution
$\delta J_j$, which depends on the site index $j$, 
and is taken from a uniform distribution: $\delta J_j \in [-\delta J, \delta J]$.
In what follows, we assume periodic boundary conditions, i.e.~$\sigma^\alpha_{N+\ell} \equiv \sigma^\alpha_\ell$,
and fix $J=\hbar = 1$. (The optimal charging time is therefore measured in units of $\hbar/J$.) We also keep the transverse field strength fixed and equal to $h=0.6$ (in units of $J$), so that the only Hamiltonian parameters that will be changed below are the strength $J_2>0$ of next-to-nearest-neighbor interactions 
and the disorder strength $\delta J>0$.
After mapping the spin-chain model into a fermionic model through a Jordan-Wigner transformation,
the next-to-nearest-neighbor interaction term can be seen to be responsible for the breakdown of integrability.

When varying the strength of $J_2$ and $\delta J$,
the disordered quantum Ising chain Hamiltonian ${\cal H} = {\cal H}_0+ {\cal H}_1$ in Eqs.~\eqref{eq:HamStatic}-\eqref{eq:HamB}
presents a rich phase diagram, which has been scrutinized in Ref.~\onlinecite{Pollmann14}. In particular, for sufficiently strong disorder, i.e.~for $\delta J > \delta J_{\rm c}$, its excitation spectrum presents a MBL phase. Conversely, for $\delta J < \delta J_{\rm c}$, the excitation spectrum presents a many-body mobility edge so that eigenstates with energy larger than a given threshold $E_0$
behave as thermal, and are believed to satisfy the {\it eigenstate thermalization hypothesis}~\cite{Polkovnikov11,Nandkishore15,DAlessio16}, while low-energy states 
are localized.
If next-to-nearest neighbor couplings are switched off ($J_2=0$), the Hamiltonian ${\cal H}$
reduces to a conventional spin-1/2 quantum Ising chain in a transverse field. In this case, any finite amount of disorder ($\delta J > 0$) in the nearest-neighbor interaction term localizes all the eigenstates, because of the one-dimensional nature of the model~\cite{Anderson58}. Setting $J_{2}=0$ yields therefore an AL phase. 

In order to evaluate the properties of the time evolved state $|\psi(\tau)\rangle = e^{-i (\mathcal{H}_0 + \mathcal{H}_1)\tau}|0\rangle$, 
it is convenient to employ a fourth-order Runge-Kutta integration of the time evolution operator~\cite{footnote3}.
Moreover, since $\{\delta J_i\}$ is a random variable, in the following we average the quantities
of interest $\mathcal{O}$ over the distribution of $\{\delta J_i\}$.
We denote by $\langle \! \langle \mathcal{O} \rangle \! \rangle$ the averaged quantity, i.e.,
\begin{equation}
\langle \! \langle \mathcal{O} \rangle \! \rangle\equiv \int \! P(\{\delta J_i\}) \mathcal{O}(\{\delta J_i\}) \; {\rm d}\{\delta J_i\}~,
\end{equation}
where $P(\{\delta J_i\})$ is the uniform probability to have a fixed value of $\delta J_i\in [-\delta J, \delta J]$.

\section{Results}
\label{sect:results}
We now analyze the charging process of our quantum Ising chain battery, focusing on the three different phases of the Hamiltonian ${\cal H}_{0}+{\cal H}_{1}$. Data shown in this section have been obtained by fixing suitable values of the parameters
$J_2$ and $\delta J$, as suggested by the phase diagram worked out in Ref.~\onlinecite{Pollmann14}. We therefore set $J_2=0$ and $\delta J =1$ to yield a representative AL phase (red color in all the figures);
$J_2=0.3$ and $\delta J=5$ to yield a MBL phase (blue color); and
$J_2=0.3$ and $\delta J=1$ to yield an ergodic phase (green color).
Of course, we have extensively tested (not shown) the dependence of all our results upon changes in the values of the parameters,
finding that the main features highlighted below are robust and independent
of the aforementioned particular choices.

\subsection{Optimal charging time}
\label{subsec:times}

\begin{figure}[t]
  \centering
  \vspace{1.em}
  \begin{overpic}[width=0.95 \columnwidth]{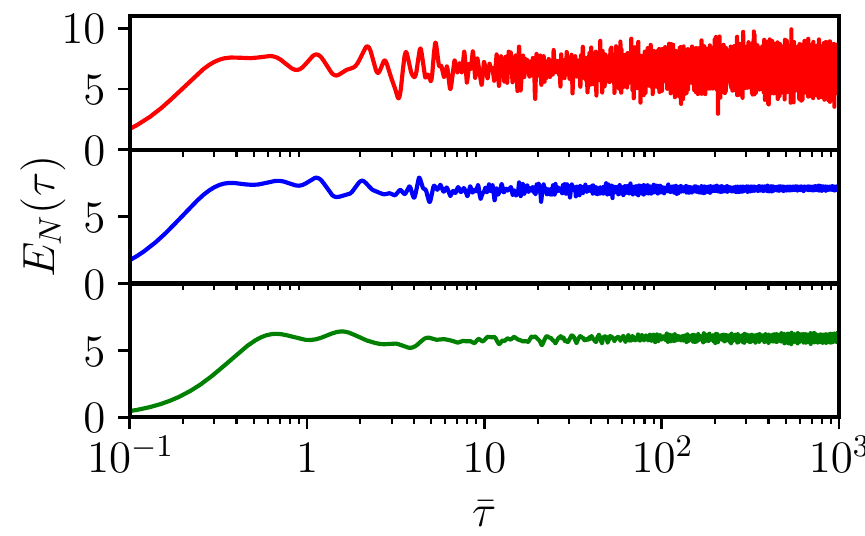}\put(0,58){\normalsize (a)}\end{overpic}\\
  \vspace{0.5 cm}
  \begin{overpic}[width=0.49\columnwidth]{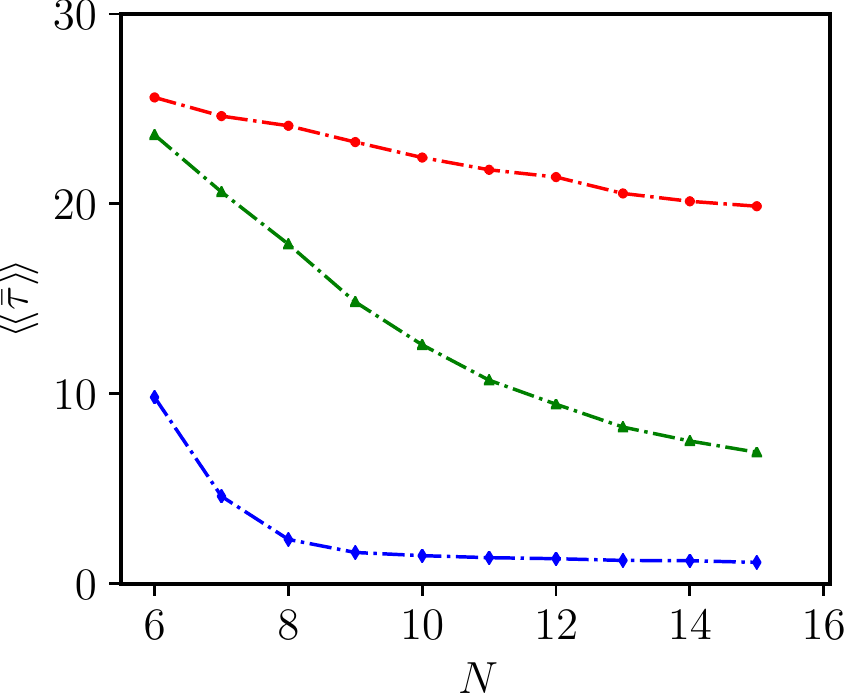}\put(4,83){\normalsize (b)}\end{overpic}
  \begin{overpic}[width=0.49\columnwidth]{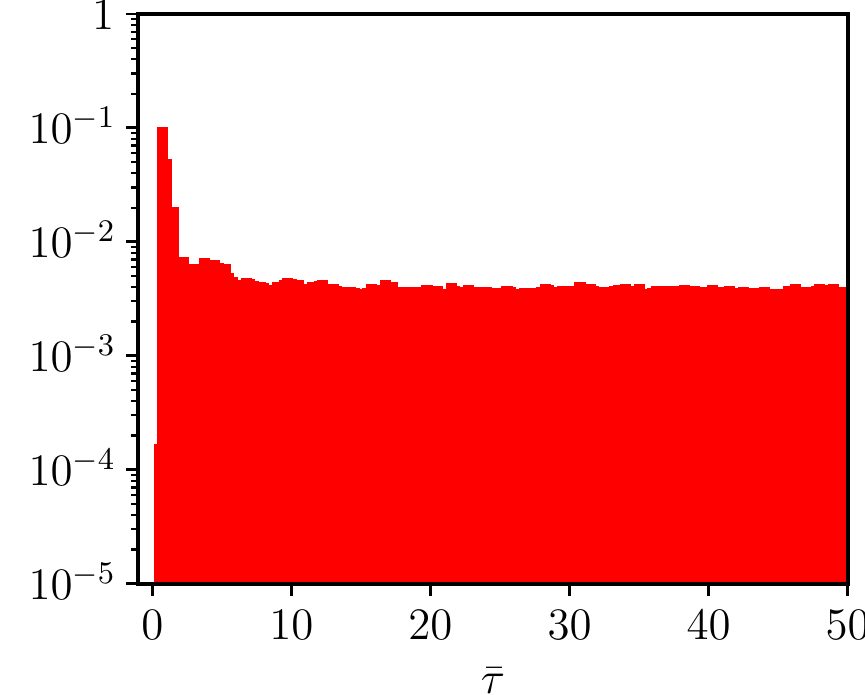}\put(4,83){\normalsize (c)}\end{overpic}\\
  \vspace*{0.2cm}
  \begin{overpic}[width=0.49\columnwidth]{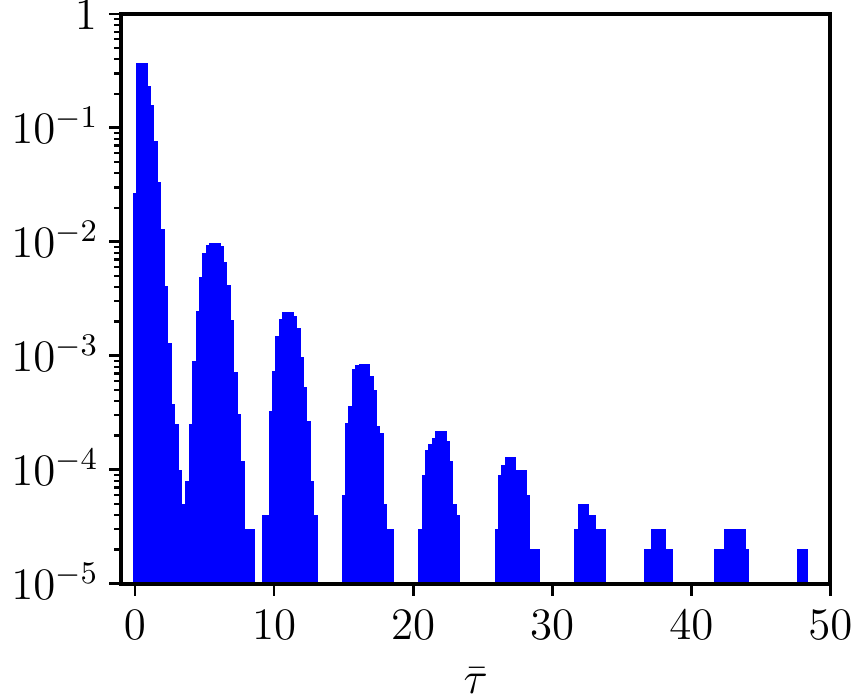}\put(4,83){\normalsize (d)}\end{overpic}
  \begin{overpic}[width=0.49 \columnwidth]{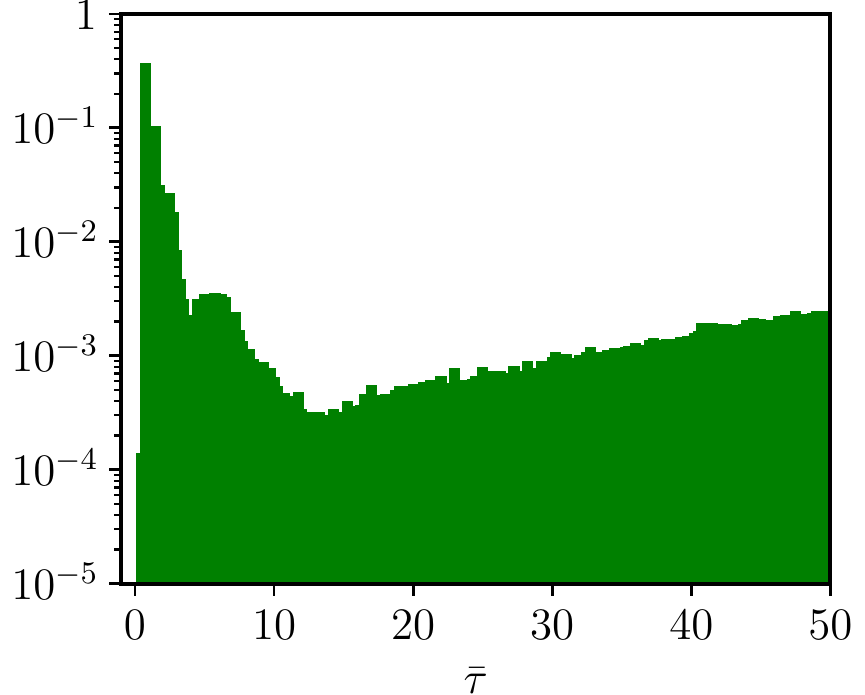}\put(4,83){\normalsize (e)}\end{overpic}
  \caption{(Color online)  
  Panel (a) The energy $E_{N}(\tau)$ as a function of the time $\tau$ for a single realization of disorder and $N=14$. The three curves correspond to different phases, according
    to the following color code that will be adopted hereafter:
    AL phase (red), MBL phase (blue), and ergodic phase (green).
    Panel (b) The optimal charging time $\langle \! \langle\bar{\tau}\rangle \! \rangle$ is plotted as a function of
    the number of units $N$ in the battery.  Data have been averaged over $10^4$ disorder
    realizations. 
    Panels (c)-(e) The probability $P(\bar{\tau})$ to find a certain value of the optimal time
    in a given phase is plotted as a function of $\bar{\tau}$.  
    Histograms in the three panels refer to a battery with $N=14$ units, in the above mentioned phases,
    and have been obtained by averaging over $10^5$ realizations. The observation time window has
    been fixed to~\cite{footnote2} $\tau_{\rm max} = 50$.}
    \label{fig:Time}
\end{figure}

First of all, the energy stored in the QB at the optimal time always scales linearly with the number
of elements in the battery, thus it is an extensive quantity irrespective of the phase of the system (see App.~\ref{sec:scaling}).
Looking more in detail at the temporal behavior of $E_N(\tau)$, we have observed consistent temporal fluctuations which have the tendency to be suppressed with increasing $N$, provided interactions $J_2$ are switched on. An example is reported in Fig.~\ref{fig:Time}(a), where we show $E_{N}(\tau)$ as a function of $\tau$, for a single realization of disorder. We note that, while the behavior at short times does not depend on the microscopic details, at longer times the AL phase is characterized by wild temporal fluctuations of the energy. These are clearly detrimental, as one would like to work with QBs that deliver energy and power in a stable manner. Interactions greatly suppress these fluctuations, yielding stable QBs. We further elaborate on temporal fluctuations in App.~\ref{sect:temporal}.

The optimal charging time $\langle \! \langle \bar{\tau} \rangle \! \rangle$, averaged over
the distribution of $\{ \delta J_j \}$, is found to decrease with $N$, see Fig.~\ref{fig:Time}(b).
Quantitative differences emerge between the various phases, with a decrease that is much faster in the MBL phase
than in the ergodic and AL phases. These results are strongly affected by the finite observation time window, especially in the AL phase where time fluctuations are most prominent and cannot be neglected. However, the fact that the optimal charging time $\langle \! \langle \bar{\tau} \rangle \! \rangle$ is shortest in the MBL phase is independent of $\tau_{\rm max}$.

More insight about the influence of the quantum many-body dynamics on the charging process can be inferred
by analyzing the full statistical distribution of the optimal times, namely the probability $P(\bar{\tau})$
to find a certain value of the optimal time in a given phase.
The results of our simulations are reported in Figs.~\ref{fig:Time}(c)-(e).
In the AL phase [panel (c)], an initial peak in such distribution is followed by a completely flat region,
which witnesses the presence of strong temporal fluctuations for any realization. $P(\bar{\tau})$ drastically changes upon switching on interactions while maintaining the system in a localized state (i.e.~in the MBL phase). A series of distinct peaks emerges, separated by forbidden regions [panel (d)].
Finally, $P(\bar{\tau})$ displays a yet different behavior in the ergodic phase [panel (e)], since in this case it exhibits
a (nearly) monotonic drop with $\bar{\tau}$. The growth of $P(\bar{\tau})$ at long times is an artifact due to the choice of a finite $\tau_{\rm max}$ (see App.~\ref{sect:dependence}). 
The statistics of the optimal times can be thus used to discriminate between the three different phases of the model, due to clear qualitative differences between the corresponding distributions $P(\bar{\tau})$.

\subsection{Extractable energy}
\label{subsec:extren}

\begin{figure*}[t]
  \centering
  \vspace{1.em}
  \begin{overpic}[width=0.68\columnwidth]{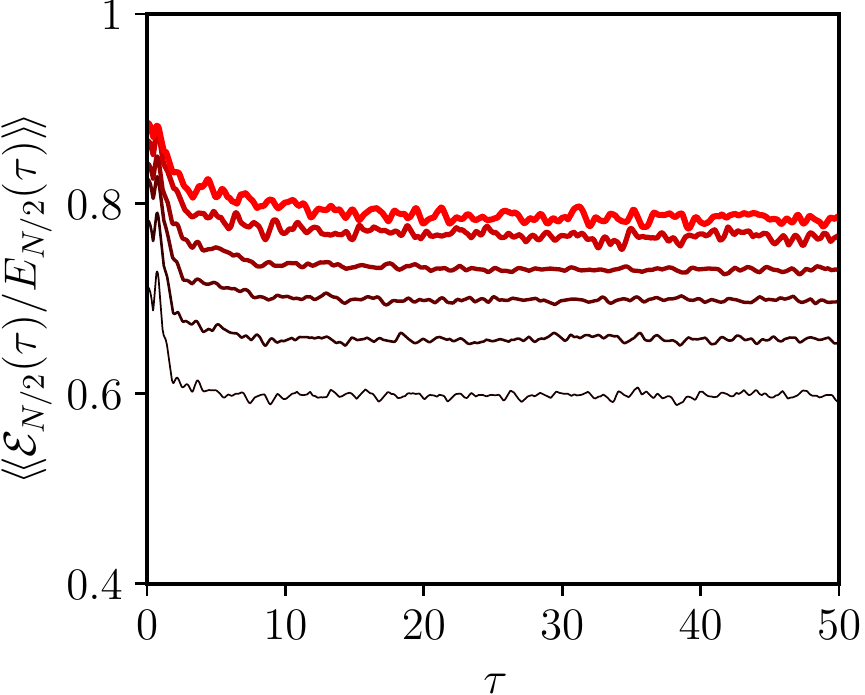}\put(4,80){\normalsize (a)}\end{overpic} 
  \begin{overpic}[width=0.68\columnwidth]{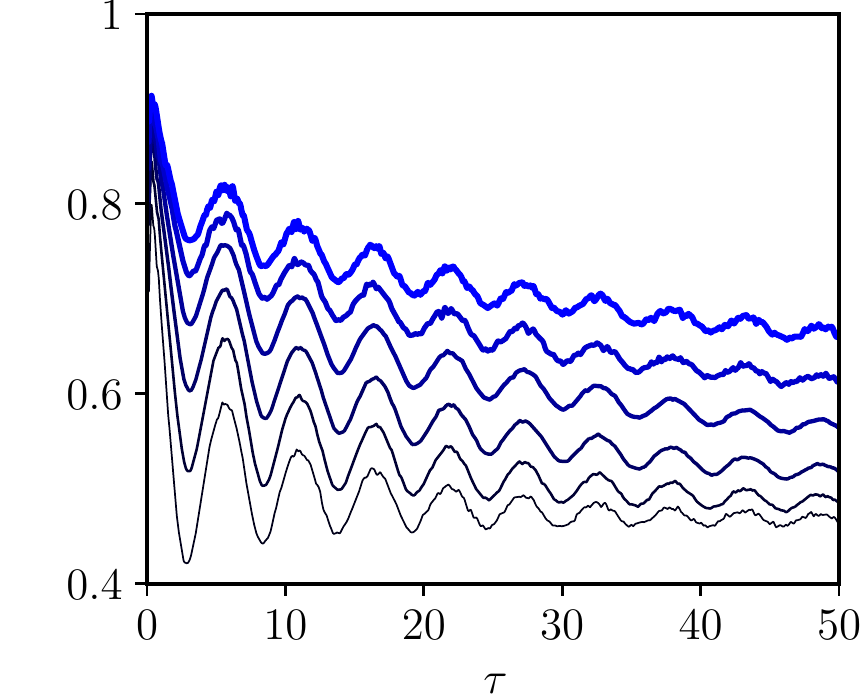}\put(4,80){\normalsize (b)}\end{overpic}
  \begin{overpic}[width=0.68\columnwidth]{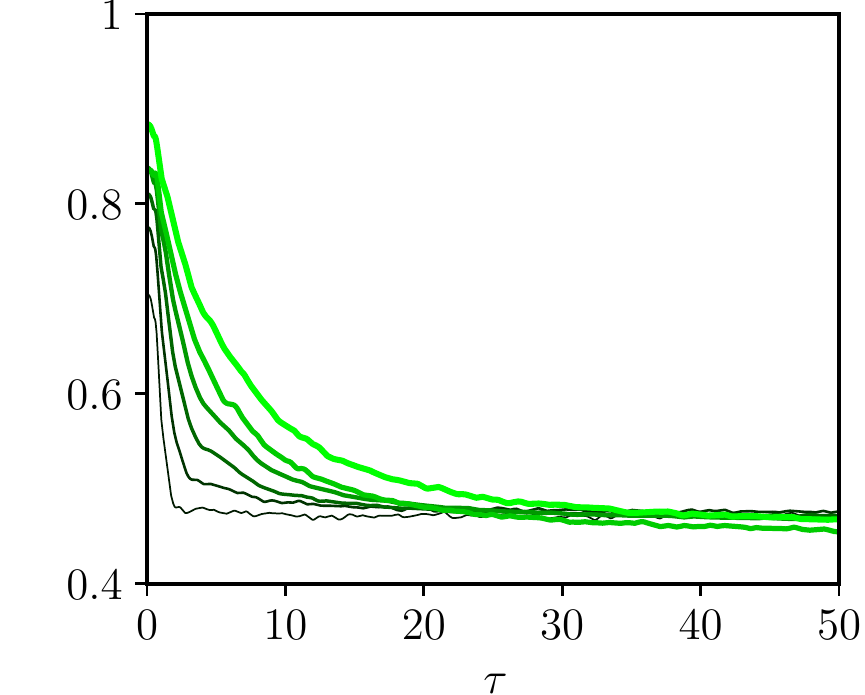}\put(4,80){\normalsize (c)}\end{overpic}\\
  \vspace*{0.2cm}
  \begin{overpic}[width=0.68\columnwidth]{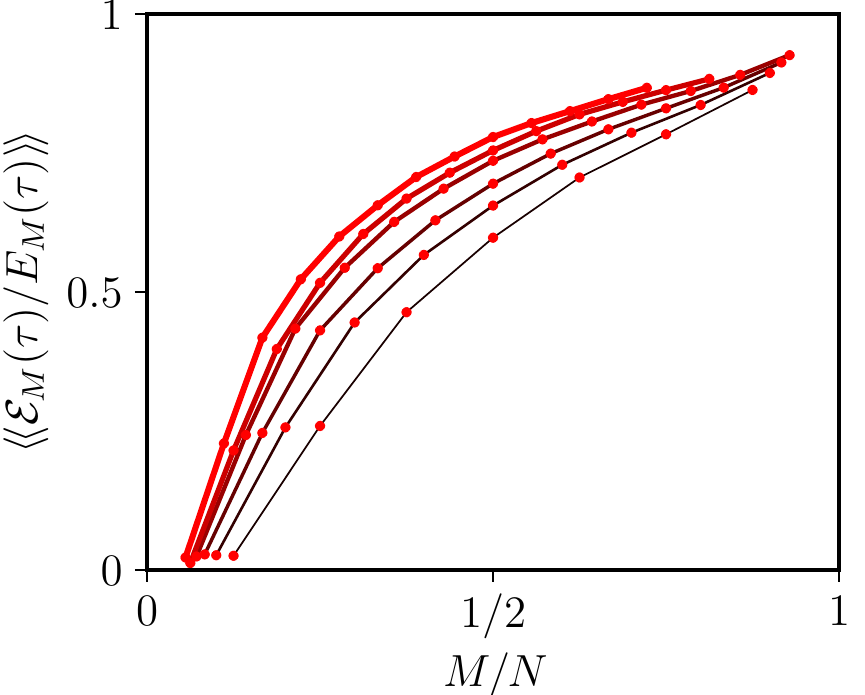}\put(4,80){\normalsize (d)}\end{overpic}
  \begin{overpic}[width=0.68\columnwidth]{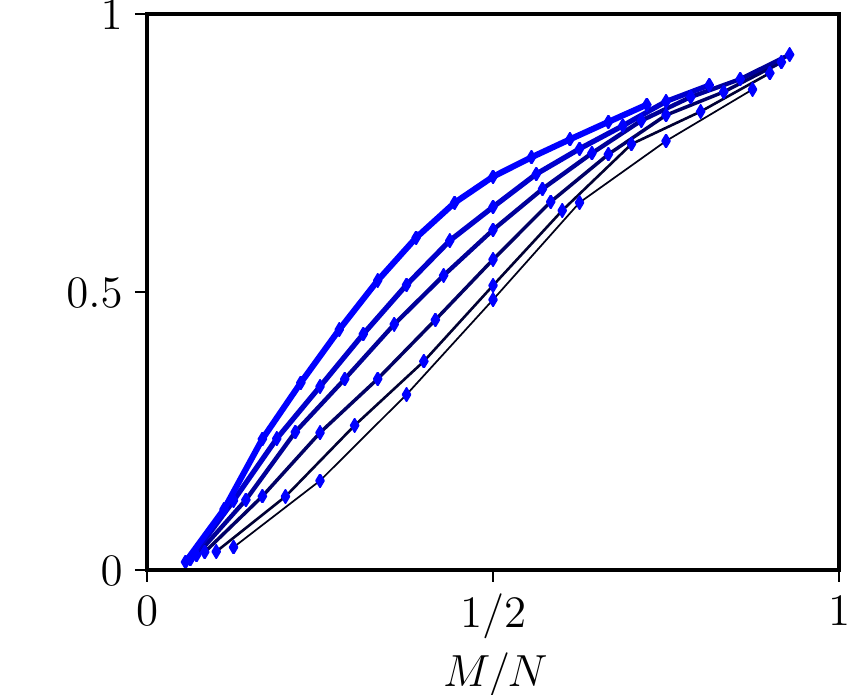}\put(4,80){\normalsize (e)}\end{overpic}
  \begin{overpic}[width=0.68\columnwidth]{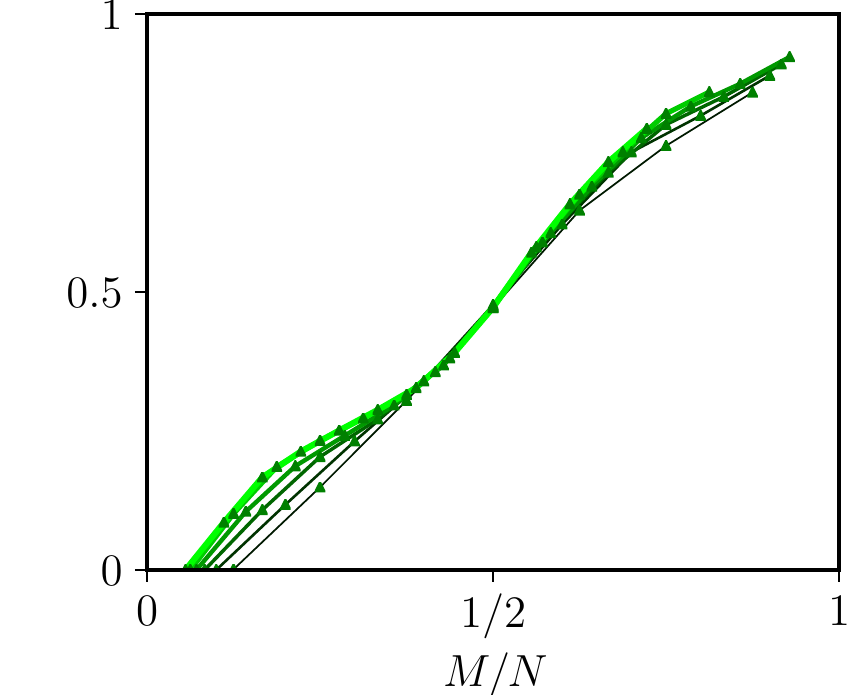}\put(4,80){\normalsize (f)}\end{overpic}
  \caption{(Color online) Panels (a,b,c)  The fraction of extractable energy
    $\langle \! \langle \mathcal{E}_{N/2}(\tau)/{E}_{N/2}(\tau) \rangle \! \rangle$ as a function of $\tau$.
    Panels (d,e,f)  The fraction of useful energy
    $\langle \! \langle \mathcal{E}_{M}(\tau)/{E}_{M}(\tau) \rangle \! \rangle$ as a function of $M/N$,
    at fixed time $\tau = 32$.
    Panels (a,d) correspond to the AL phase, panels (b,e) to the MBL phase, and panels (c,f) to the ergodic phase. 
    The various solid lines correspond to different values of the number $N$ of quantum cells: $N = 8,10,12,14,16,18$.
    The lines thicken and brighten with increasing $N$.
    Data are averaged over $10^{3}$ disorder realizations for $N$ up to $14$,
    and over $10^2$ realizations in the cases $N=16$ and $18$.}
  \label{fig:ERG}
\end{figure*}

We now focus on the amount of extractable energy $\mathcal{E}_M(\tau)$. 
In particular, we are interested in: (i) the disorder-averaged fraction of useful energy in half of the battery,
i.e.~$\langle \! \langle \mathcal{E}_{N/2}(\tau)/{E}_{N/2}(\tau) \rangle \! \rangle$ as a function of $\tau$; (ii) the same fraction at fixed $\tau$, viewed as a function of the subsystem size $M \leq N$,
i.e.~$ \langle \! \langle \mathcal{E}_M(\tau)/{E}_M(\tau) \rangle \! \rangle$. 
These quantities are of interest because performing operations on all qubits may be experimentally challenging, while it is reasonable to assume fully global control only on a restricted number $M$ of battery unities.
These two quantities are reported in Fig.~\ref{fig:ERG}, upper and lower panels, respectively.
Results in the AL phase [panels (a,d)], are compatible with the following asymptotics~\cite{Andolina19}:
\begin{eqnarray}
  \lim_{N\to\infty}\langle \! \langle\mathcal{E}_{N/2}(\tau)/{E}_{N/2}(\tau) \rangle \! \rangle & = & 1~, \label{eq:asymp1}\\
  \lim_{M \to \infty} \langle \! \langle\mathcal{E}_{M}(\tau)/{E}_{M}(\tau) \rangle \! \rangle & = & 1~.   \label{eq:asymp2}
\end{eqnarray}
Indeed, due to the integrability of the model at $J_{2}=0$, only a small portion of the entire Hilbert space is visited during the quantum dynamics.
Consequently, the difference between energy and ergotropy
is expected to become negligible in the thermodynamic limit~\cite{Andolina19}, in accordance with the trends found  numerically. In the MBL phase [panels (b,e)], the presence of an extensive number of exponentially localized
constants of motion~\cite{Abanin18,Nandkishore15} keeps the system's dynamics ``frozen" as well, and the bipartite entanglement entropy
fulfills an area-law scaling (up to logarithmic corrections)~\cite{Abanin18,Nandkishore15}.
Therefore a scenario qualitatively analogous to the AL phase occurs~\cite{Andolina19}.
We checked that the same conclusions apply for simpler integrable non-disordered systems (see App.~\ref{sect:ergotropy}).

A completely different situation occurs in the ergodic phase [panels (c,f)].
Indeed, the behavior of the fraction of extractable energy with $\tau$ suggests that
$\langle \! \langle \mathcal{E}_{N/2}(\tau)/{E}_{N/2}(\tau) \rangle \! \rangle$ saturates
to a finite constant when the thermodynamic limit is taken.
This is a signature of ergodicity. All the Hilbert space is visited
during the dynamical evolution, and the evolved state locally resembles a canonical ensemble. 
Hence, the arguments of Ref.~\onlinecite{Andolina19} do not hold in this case.
Similar conclusions apply to generic non-integrable and non-disordered systems (see App.~\ref{sect:ergotropy}).
We note that $\langle \! \langle \mathcal{E}_{M}(\tau)/{E}_{M}(\tau) \rangle \! \rangle$ 
tends to form a plateau in the region $M/N < 1/2$, which is reminiscent of the redundancy plateau
emerging in the mutual information as a function of $M/N$ witnessing quantum Darwinism~\cite{Zurek_09, Riedel_12}.

\section{Discussion and conclusions}
\label{sect:discussion}

The results presented in this work are not qualitatively affected by the specific choice of the Hamiltonian
in Eqs.~\eqref{eq:HamStatic}-\eqref{eq:HamB}. We have performed numerical simulations (see App.~\ref{sect:heisenberg}) to verify
that similar conclusions apply for a QB described by a XXZ Heisenberg spin chain in the presence of a random field
along the $\hat{\bm z}$ axis~\cite{Oganesyan07,Znidaric08,Pal10, Laflorencie15}. 

Our findings are amenable to experimental verification. Interacting spin-chain Hamiltonians can be implemented, e.g., by using trapped ions~\cite{Abanin18,Smith16,Maier19}. In Ref.~\onlinecite{Maier19}, for example, energy transport has been studied in a trapped-ion setup containing $N=10$ spins, in the presence of static disorder and dephasing noise, paving the way for studies of energy flow in MBL spin-chain-based QBs. Exact numerical simulations for relatively small values of $N$, as those carried out in this work, are therefore relevant for interpreting experiments in trapped-ion set-ups and other set-ups simulating a variety of interacting-qubit models, such as superconducting circuits~\cite{Xu18,Roushan17}.

In the future it will be interesting to study the role of interactions and disorder in the more general context of thermal nano-machines and quantum thermodynamics~\cite{Campisi11,Gelbwaser-Klimovsky15,Goold16,Vinjanampathy16,Strasberg17,Niedenzu18,Halpern19,Watanabe19}.

During the completion of this work we became aware of a recent interesting analysis of disordered QBs~\cite{SenDe}. The authors of this work, however, did not study the role of many-body localization on the figures of merit.

\acknowledgments
We wish to thank V. Giovannetti and E. Vicari for useful discussions.

\appendix

\section{Scaling of the stored energy and the average power with $N$}
\label{sec:scaling}

\begin{figure}[!t]
  \begin{overpic}[width=0.8\columnwidth]{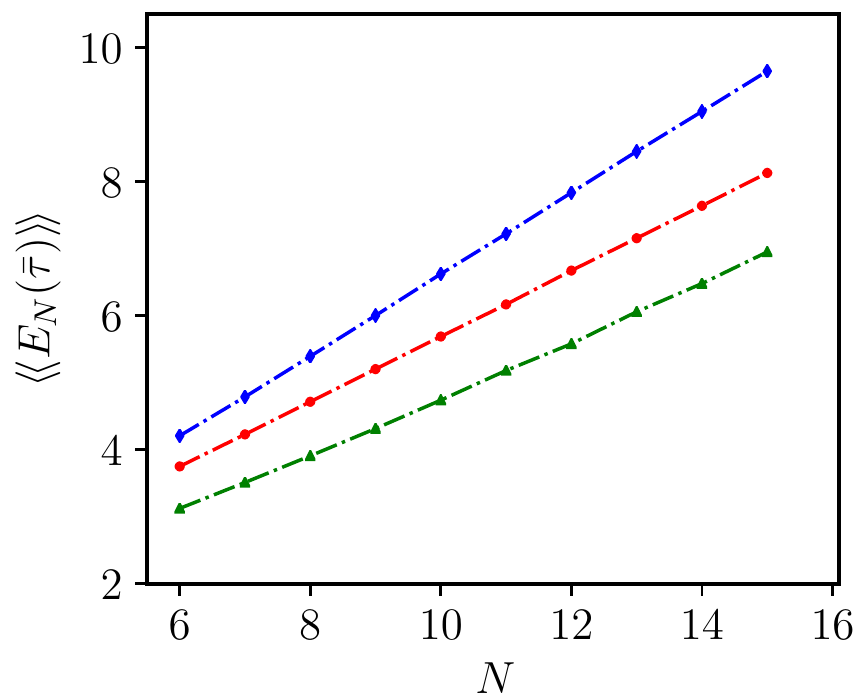}\put(4,80){(a)}\end{overpic}
  \begin{overpic}[width=0.8\columnwidth]{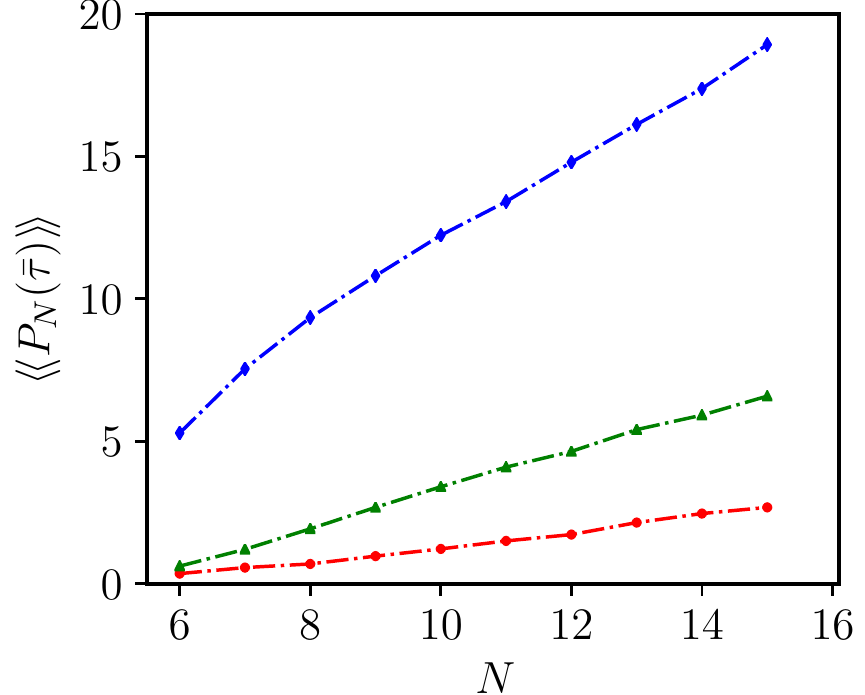}\put(4,80){(b)}\end{overpic}
  \caption{(Color online) Panel (a) The energy stored in the QB at the optimal time $\langle \! \langle E_{N} (\bar{\tau})\rangle \! \rangle$, as a function of $N$, in the AL phase (red), MBL phase (blue), and ergodic phase (green). Panel (b) The average power in the QB at the optimal time $\langle \! \langle P_{N} (\bar{\tau})\rangle \! \rangle$, as a function of $N$. Same color code as in panel (a). Data have been obtained by averaging over $10^4$ realizations.} 
  \label{fig:scaling}
\end{figure}

In order to ensure that the mean local energy~\eqref{eq:stored energy} stored in the QB at the optimal time is an extensive quantity and is thus well defined in the thermodynamic limit, we have verified that it increases
linearly with the number of quantum cells $N$:
\begin{equation}
  \langle \! \langle E_N(\bar{\tau}) \rangle \! \rangle \sim N~.
\end{equation}
This is shown in panel (a) of Fig.~\ref{fig:scaling}, where the various curves refer to the three different phases introduced in Sect.~\ref{sect:results}.
We observe that the injected energy is generally smaller in the ergodic phase, thus confirming the fact
that localized phases yield the optimal scenario where a QB can be operated. Indeed, in these phases, better performances can be achieved, both in terms of charging energy
and of work extraction.

More in detail, we have found numerical evidence that increasing both the disorder strength $\delta J$
and the next-to-nearest neighbor interactions, generally leads to an amplification of the energy stored in the QB, for fixed $N$.
Indeed, while for the three representative cases reported in the figure the MBL phase appears as the one which
optimizes the amount of stored energy, for a given value of $N$, this observation depends
on the choice of $\delta J$. (We remind the reader that, as discussed in Sect.~\ref{sect:results}, data for the AL phase
correspond to $\delta J = 1$, while those for the MBL phase correspond to $\delta J = 5$.) Choosing the same value of $\delta J$ for both cases would result in a slightly larger amount of injected
energy ($\approx 15\%$) in the AL phase, despite the presence of detrimental wild temporal fluctuations
[see Fig.~\ref{fig:Time}(a) and App.~\ref{sect:dependence}].
However, one can show that the average value of the injected energy, defined in Eq.~\eqref{eq:Ener_avg} below,
is the same for the two phases at fixed $\delta J$.

It is also worth mentioning that even the total energy $E^{\rm tot}_N(t)$ of the time evolved state $|\psi(t) \rangle$
scales linearly with $N$. Indeed, as already commented in Sect.~\ref{subsec:QB}, such quantity remains constant
at all times, except for the switching points $t=0$ and $t=\tau$.
Therefore
\begin{equation}
  E^{\rm tot}_N(\tau) = \langle \psi(0) |\mathcal{H}(0^+)| \psi(0) \rangle = \langle 0 |\mathcal{H}_0 + \mathcal{H}_1| 0 \rangle~.
\end{equation}
Since $|0\rangle$ is the ground state of $\mathcal{H}_0=\sum_{j=1}^N {h}_j$, we have that
$\langle 0 | \mathcal{H}_1| 0 \rangle = 0$, and $\langle 0 | \mathcal{H}_0| 0 \rangle = 0$. 

Furthermore, we checked numerically (not shown) that the variance of the total Hamiltonian $\mathcal{H}(0^+)=\mathcal{H}_0 + \mathcal{H}_1$ scales linearly with $N$, namely:
\begin{equation}
  \langle \psi(0) |\mathcal{H}^2(0^+)| \psi(0) \rangle = \langle 0 | \mathcal{H}^2_1| 0 \rangle \sim N~,
\end{equation} 
where we used the fact $ \mathcal{H}_0| 0 \rangle=0 $. The fact that the mean and the variance scale extensively with $N$ ensures that the model is well defined in the thermodynamic limit.

Another relevant figure of merit for a QB is the average charging power, namely
\begin{equation}
  \langle \! \langle P_N(\bar{\tau}) \rangle \! \rangle \equiv \langle \! \langle E_N(\bar{\tau}) /  \bar{\tau} \rangle \! \rangle ~.
\end{equation}
As the average time $\langle \! \langle \bar{\tau} \rangle \! \rangle$, this quantity measures the speed of the charging process. 
The average charging power is shown in panel (b) of Fig.~\ref{fig:scaling}, which displays a linear behavior for large enough
values of $N$, which is independent of the particular quantum phase.
It is interesting to note that, even with respect to this figure of merit, the MBL phase proves
to be quantitatively optimal.

\section{Temporal fluctuations}
\label{sect:temporal}

Any energy storage device is expected to deliver energy and power in a stable fashion. We have investigated the stability of the charging process by analyzing the temporal fluctuations of the mean local energy $E_N(\tau)$.
These can be quantified by means of the time-averaged variance
\begin{equation}
  \overline{\delta E_N^2}= \frac{1}{\tau_{\rm max}} \int_0^{\tau_{\rm max}} d\tau\Big( E_N(\tau) -\overline{E_N} \Big)^2 ~,
\end{equation}
where $\tau_{\rm max}$ denotes a given observation time window and $\overline{E_N}$ stands for the average value
of the energy of the battery, in the time interval $\tau \in [0,\tau_{\rm max}]$:
\begin{equation}
  \overline{E_N}  = \frac{1}{\tau_{\rm max}} \int_0^{\tau_{\rm max}} d\tau E_N(\tau)~.
  \label{eq:Ener_avg}
\end{equation}
 In the cases in which disorder is present ($ \delta J\neq0$), we have also performed averages of the above quantities over a number of different disorder realizations, thus obtaining the averaged quantities $\langle \! \langle\overline{\delta E_N^2}\rangle \! \rangle$ and $\langle \! \langle\overline{E_N}\rangle \! \rangle$.

\begin{figure}[!t]
  \centering
  \begin{overpic}[width=0.85\columnwidth]{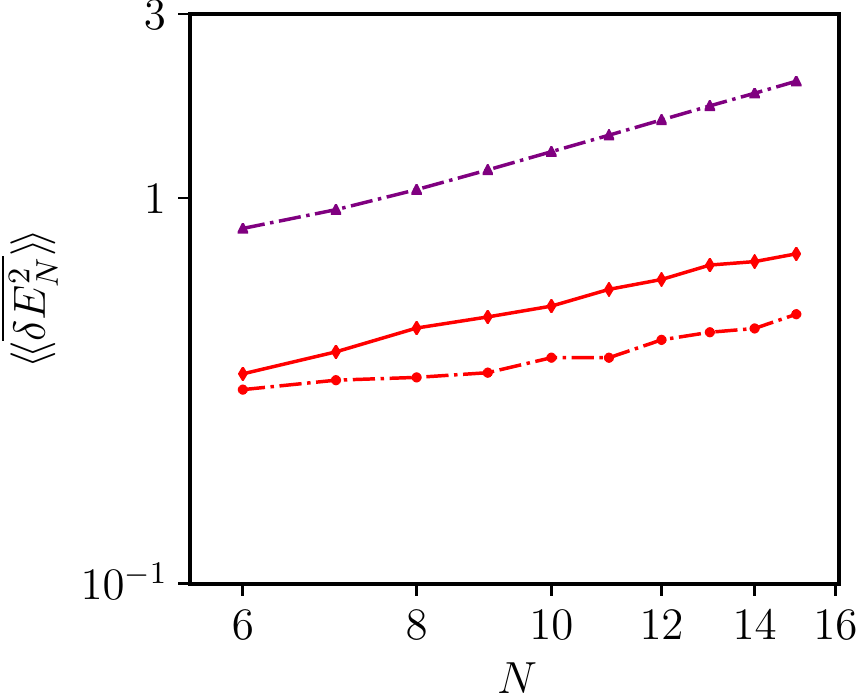}\put(4,80){\normalsize (a)}\end{overpic}
  \begin{overpic}[width=0.85\columnwidth]{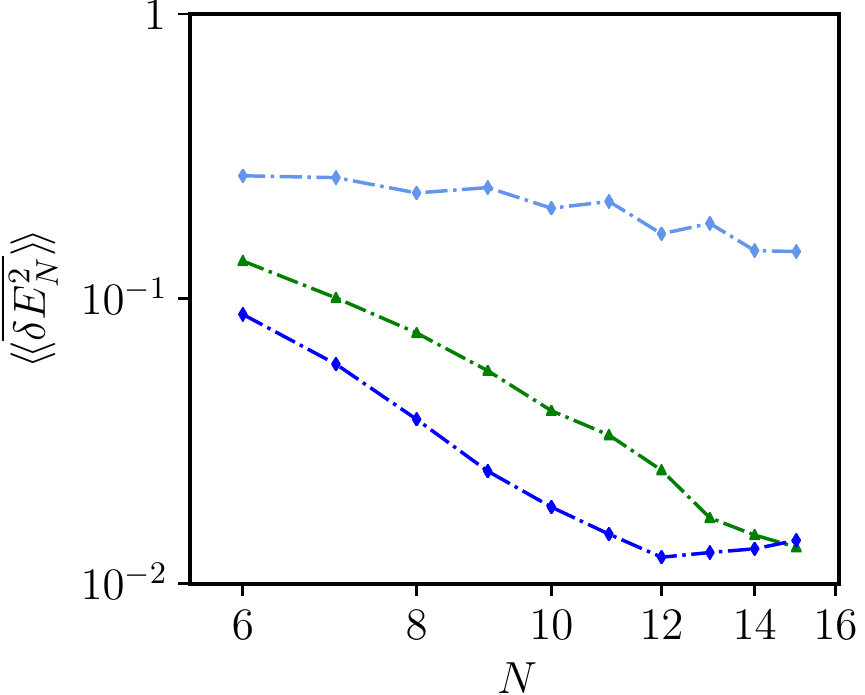}\put(4,80){\normalsize (b)}\end{overpic}
  \caption{(Color online) The time average of the variance of the mean local energy 
    as a function of time, $\overline{\delta E_N^2}$, averaged over $10^2$ realizations and
    with an observation time window $\tau_{\rm max} = 10^3$, as a function of $N$.
    Panel (a) describes non-interacting situations ($J_2 = 0$), including the AL phase (with 
    $\delta J = 1$ or $5$, corresponding to red lines with circles or diamonds respectively)
    and the clean integrable model ($\delta J = 0$, purple line).
    Panel (b) describes interacting situations ($J_2 = 0.3$), including the ergodic phase ($\delta J = 1$, green line),
    the MBL phase ($\delta J = 5$, blue line), and the clean non-integrable model ($\delta J = 0$, cyan line).}
  \label{fig:FLUCTUATION}
\end{figure}

The outcomes of our numerical simulations for the averaged temporal fluctuations $\langle \! \langle\overline{\delta E_N^2}\rangle \! \rangle$
are reported in Fig.~\ref{fig:FLUCTUATION}. We separately discuss the cases $J_{2}=0$ and $J_{2}\neq 0$
[panels (a) and (b), respectively], since they display different qualitative behaviors.
Panel (a) shows the data corresponding to two situations within the AL phase (red data sets, with $\delta J =1$ or $5$),
compared with the clean (i.e., non-disordered) non-interacting case (purple data set).
We observe that fluctuations always increase with the number $N$ of quantum cells, thus mining the
usefulness of any QB operating in this regime.
The trend is compatible with a large-$N$ linear growth, even though we have not further investigated this issue.

Panel (b) focuses on prototypical interacting situations for $J_2=0.3$, comparing the data for the MBL phase (blue)
with those for the ergodic (green) and the clean interacting case (cyan data set).
In all such cases we observe a general decrease of fluctuations with $N$, in stark contrast with the non-interacting
situations of panel (a). In terms of magnitudes, the MBL phase results in smaller fluctuations at fixed $N$,
thus representing the optimal regime at which a disordered QB should be operated, from a stability point of view.

\section{Dependence of $P(\bar{\tau})$ on the observation time window}
\label{sect:dependence}

The analysis of the optimal charging times for the QB, reported in Fig.~\ref{fig:Time},
has been performed for a fixed observation time window, i.e.~$\tau_{\rm max} = 50$.
This may have some influence on the obtained results, since it introduces an artificial temporal cutoff.
We have indeed checked that, if temporal fluctuations of the local energy $E_N(\tau)$ are not suppressed
in time nor in $N$, such cutoff becomes particularly important.
Specifically, the presence of next-to-nearest neighbor spin-spin interactions is seen to systematically suppress
time fluctuations with $N$, while this is not the case for $J_2=0$
(i.e., in the AL phase, where they are particularly pronounced).
As a consequence, enlarging the observation time window would cause a consistent increase of the measured optimal
charging time in the AL phase. The corresponding histogram would nevertheless remain qualitatively
unaffected, with a characteristic long and flat tail extending at arbitrarily large values of $\bar{\tau}$
[see Fig.~\ref{fig:Time}(c)].

\begin{figure}[!t]
  \begin{overpic}[width=1\columnwidth]{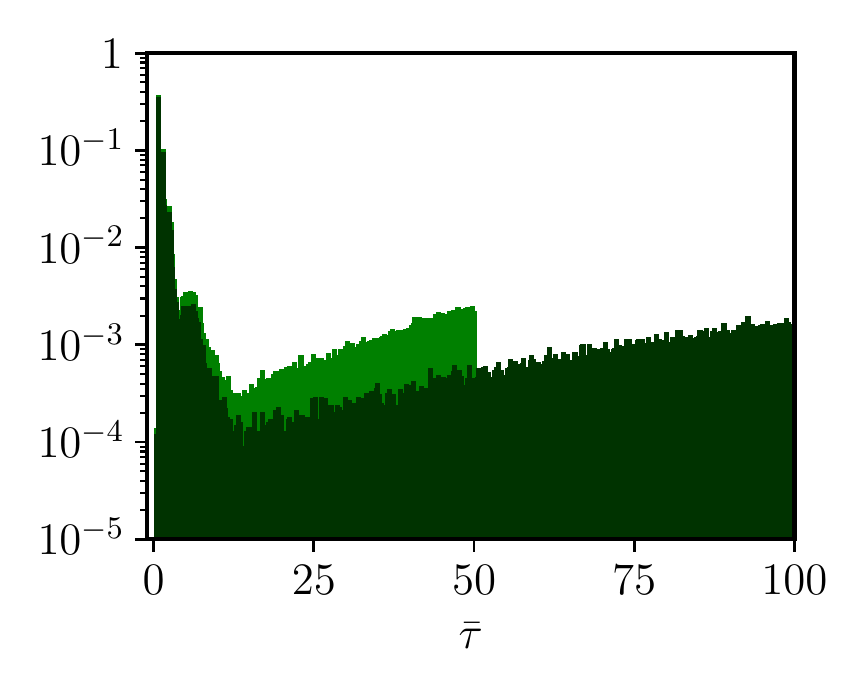}\put(4,80){}\end{overpic}
  \caption{Probability $P(\bar{\tau})$ in the ergodic phase as a function of $\bar{\tau}$, for two fixed values of $\tau_{\rm max}=50$ (green histogram)
    and $\tau_{\rm max}=100$ (dark green histogram). All other parameters are the same as in Fig.~\ref{fig:Time} panel (e),
    the representative case for the ergodic phase. In order to calculate the distribution, we simulated
    $10^5$ realizations of disorder.}
  \label{fig:TimeS}
\end{figure}

A less pronounced influence of $\tau_{\rm max}$ on $\bar{\tau}$ can be witnessed in the ergodic phase ($J_2 \neq 0$),
since time fluctuations are indeed suppressed with $N$.
However, the shape of $P(\bar{\tau})$, shown in Fig.~\ref{fig:Time}(e), may be significantly 
modify in its tail.
This is shown in Fig.~\ref{fig:TimeS}, where we compare the statistical distribution of optimal times for two different values of $\tau_{\rm max}$, i.e.~$\tau_{\rm max} = 50$ (light green) and $\tau_{\rm max} = 100$ (dark green).
It is clearly visible that, while the distribution for small values of $\bar{\tau}$ is insensitive to the choice
of $\tau_{\rm max}$, the late-time growth for $\bar{\tau}\sim \tau_{\rm max}$ is a finite-time window
effect, which depends on the particular choice of $\tau_{\rm max}$.

Finally, the charging time behavior in the MBL phase is less affected by $\tau_{\rm max}$,
as, for a typical disorder realization, time fluctuations are smaller and $\bar{\tau} \ll \tau_{\rm max}$.

%
\section{Ergotropy for a non-disordered spin-chain QB}
\label{sect:ergotropy}

We now discuss the extractable work in a QB where disorder is absent ($\delta J=0$),
focusing on the two paradigmatic cases where interactions are either absent ($J_2=0$, purple) 
or present ($J_2=0.3$, cyan).
For the sake of clarity, we keep the other parameters fixed to 
the same values as in Sect.~\ref{sect:results}, i.e.~$J=1$ and $h=0.6$. The considerations reported below are not qualitatively affected by this choice.
We also remind the reader that, in this case, statistical averages are not required, since the model is completely deterministic.

The spin-chain Hamiltonian $\mathcal{H} = \mathcal{H}_0 + \mathcal{H}_1$
[see Eqs.~\eqref{eq:HamStatic} and~\eqref{eq:HamB}], with $\delta J = 0$
and $J_2=0$, reduces to the integrable quantum Ising chain in a transverse field.
Such model admits a simple and manageable solution in terms of free quasiparticles,
after first mapping it into a quadratic fermion model through a Jordan-Wigner
transformation, and then performing a Bogoliubov rotation~\cite{Sachdev}.
On the other hand, for $\delta J=0$ and $J_2\neq 0$, it turns out to be non-integrable,
since the Jordan-Wigner transformation maps the next-to-nearest neighbor spin-spin coupling term
($\sigma^x_j \sigma^x_{j+2}$) into a quartic fermion operator.

\begin{figure}[!t]
  \begin{center}
  \begin{overpic}[width=0.49\columnwidth]{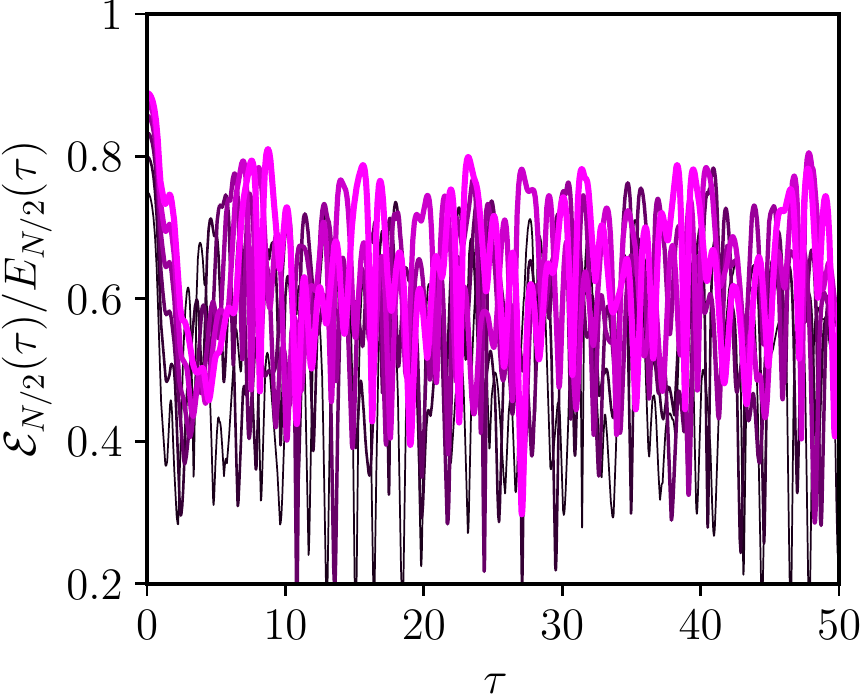}\put(2,84){\normalsize (a)}\end{overpic}
  \begin{overpic}[width=0.49\columnwidth]{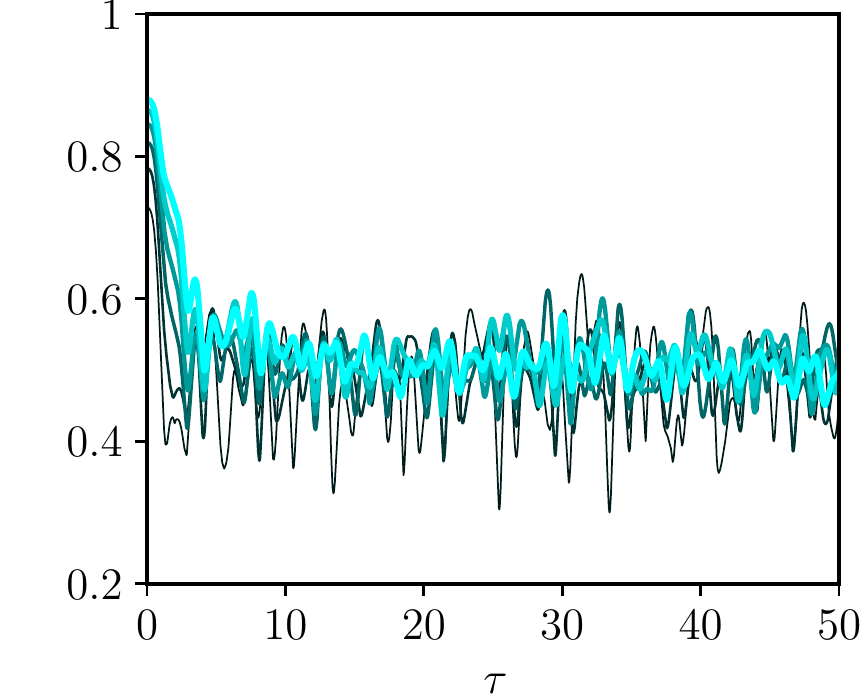}\put(2,84){\normalsize (b)}\end{overpic}\\
  \vspace{0.6cm}
  \begin{overpic}[width=0.49\columnwidth]{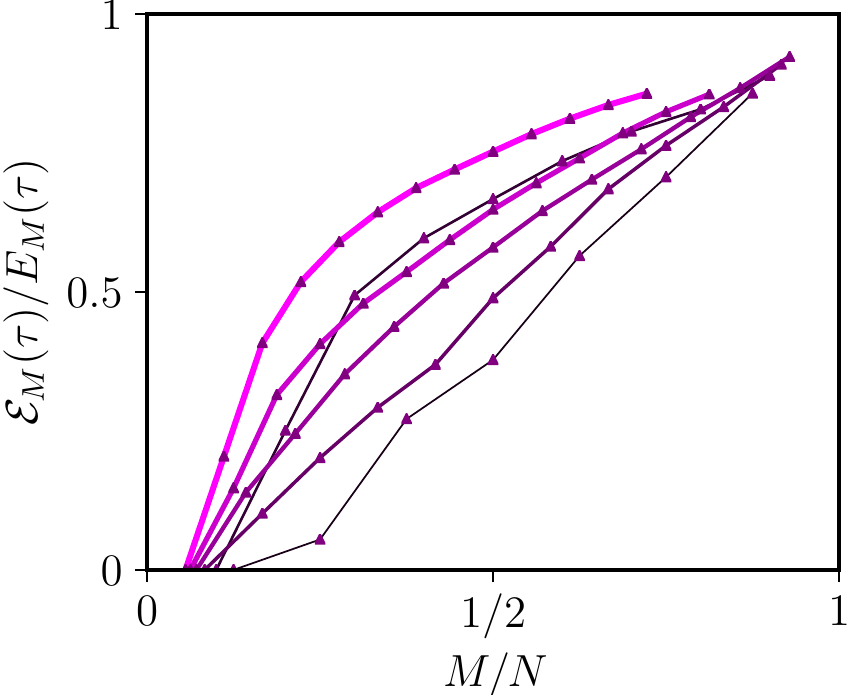}\put(2,84){\normalsize (c)}\end{overpic}
  \begin{overpic}[width=0.49\columnwidth]{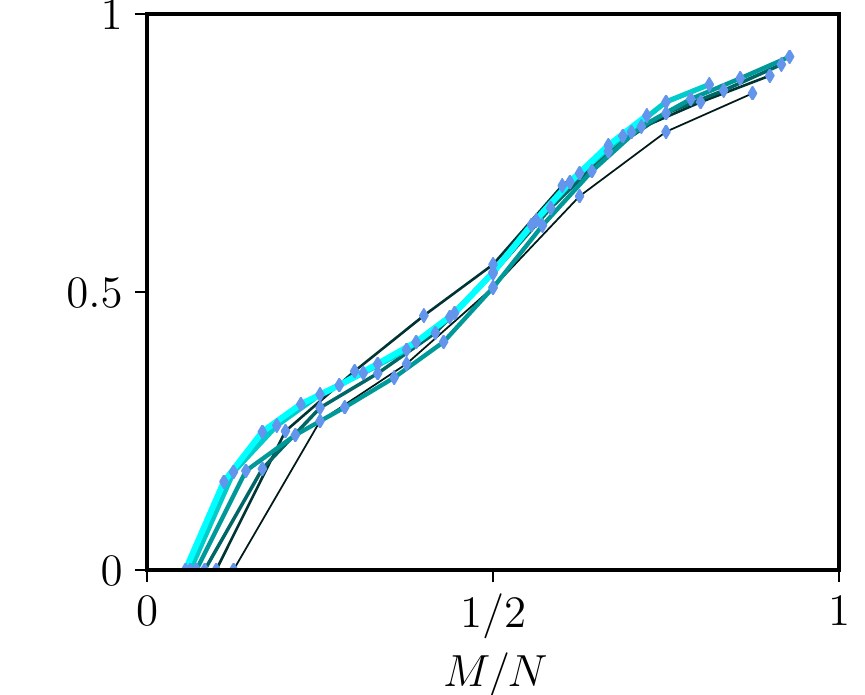}\put(2,84){\normalsize (d)}\end{overpic}
    \end{center}
  \caption{(Color online) Analysis of the extractable work for a QB modeled by a clean spin chain ($\delta J = 0$).
    Panels (a,b) show the fraction of useful energy $\mathcal{E}_{N/2}(\tau)/{E}_{N/2}(\tau)$
    as a function of $\tau$. Panels (c,d) show the fraction of useful energy
    $\mathcal{E}_{M}(\tau)/{E}_{M}(\tau)$ as a function of the fraction $M/N$ at the given time $\tau=32$. Panels (a,c) [respectively, Panels (b,d)] correspond to the non-interacting (respectively, interacting) phase,
    which is shown in purple (respectively, cyan). Different solid lines correspond to different values
    of the number of cells in the QB: $N = 8,10, \ldots, 18$.
    An increase in thickness and brightness of the various lines corresponds to an increase of $N$.}
  \label{fig:ERGS}
\end{figure}

We start by focusing on the non-interacting case (quantum Ising chain in a transverse field).
Panel (a) of Fig.~\ref{fig:ERGS} displays the fraction of useful energy for half of the system
$\mathcal{E}_{N/2}(\tau)/{E}_{N/2}(\tau)$ as a function of $\tau$, while panel (c) displays the same fraction
as a function of the fraction $M/N$, i.e. $\mathcal{E}_{M}(\tau)/{E}_{M}(\tau)$, at the given time $\tau=32$. 
We first recognize the presence of strong oscillations in time, reminiscent of the wild temporal fluctuations for $J_2=0$ (see above).
Moreover, the trend with $N$ suggests that $\mathcal{E}_{N/2}(\tau)/{E}_{N/2}(\tau) \to 1$
and $ \mathcal{E}_{M}(\tau)/{E}_{M}(\tau) \to 1$, analogously to what has been reported in the AL and MBL phases,
cf., Eqs.~\eqref{eq:asymp1}-\eqref{eq:asymp2} [see panels (a,d) and (b,e) of Fig.~\ref{fig:ERG}, respectively].
Indeed, as it occurs in a localized phase, due to the presence of multiple constants of motion
(i.e. the number operators corresponding to the fermionic quasiparticles which diagonalize the model),
the quantum dynamics is constrained in a small subportion of the full Hilbert space, and the difference between energy
and ergotropy is expected to become negligible with increasing $N$.

Panels (b,d) of Fig.~\ref{fig:ERGS} show the same quantities of panels (a,c), but
for the interacting case (quantum Ising chain in a transverse field, with next-to-nearest neighbor couplings).
The behavior of the useful energy versus time and the fraction of the system size
suggests that it saturates to a finite constant (different from one), when the thermodynamic limit is taken.
Again, we can ascribe this fact to the presence of interactions (i.e.~$J_{2}$),
which are known to break integrability and thus to induce ergodicity, similarly to what has been reported
in the ergodic phase for the disordered QB model [panels (c,f) of Fig.~\ref{fig:ERG}].
We conclude that the asymptotic behavior of the ergotropy
is strongly connected with the integrability of the underlying dynamical model.

\section{Disordered Heisenberg spin-chain model}
\label{sect:heisenberg}

In order to verify the robustness of our findings and claims,
we have also performed numerical simulations for a different spin-chain QB.
Specifically, we have considered an alternative QB model which undergoes
the same charging protocol described in Sect.~\ref{subsec:QB}, and is characterized
by a time-dependent Hamiltonian of the form as in Eq.~(1).
The free Hamiltonian $\mathcal{\tilde H}_0$ is constituted by single-particle spin-flip terms
\begin{eqnarray}
  \mathcal{\tilde H}_0 &=& \frac{J}{2} \sum_j \big( \sigma^x_j \sigma^x_{j+1} +
  \sigma^y_j \sigma^y_{j+1} \big) \nonumber\\
  &=&
  J \sum_j \big( \sigma^+_j \sigma^-_{j+1} + {\rm h.c.} \big)~,
\end{eqnarray}
where $\sigma^\pm_j = \tfrac12 (\sigma^x_j \pm \sigma^y_j)$
denotes raising/lowering operators for spin-1/2 particles.
The charging part $\mathcal{\tilde H}_1$ is the sum
of a nearest-neighbor spin-spin coupling along the $\hat{\bm z}$ axis
(namely, the interaction), and a random field along the same direction:
\begin{equation}
  \mathcal{\tilde H}_1 = \frac{\Delta}{2} \sum_j \sigma^z_j \sigma^z_{j+1} +
  \sum_j h^z_j \sigma^z_j~,
\end{equation}
where $\Delta$ denotes the strength of the $\hat{\bm z}$-axis anisotropy
and $h^z_j$ is a randomly-varying field strength,
which is taken from a uniform distribution $h^z_j \in [-W, W]$.

\begin{figure}[t]
  \begin{overpic}[width=1\columnwidth]{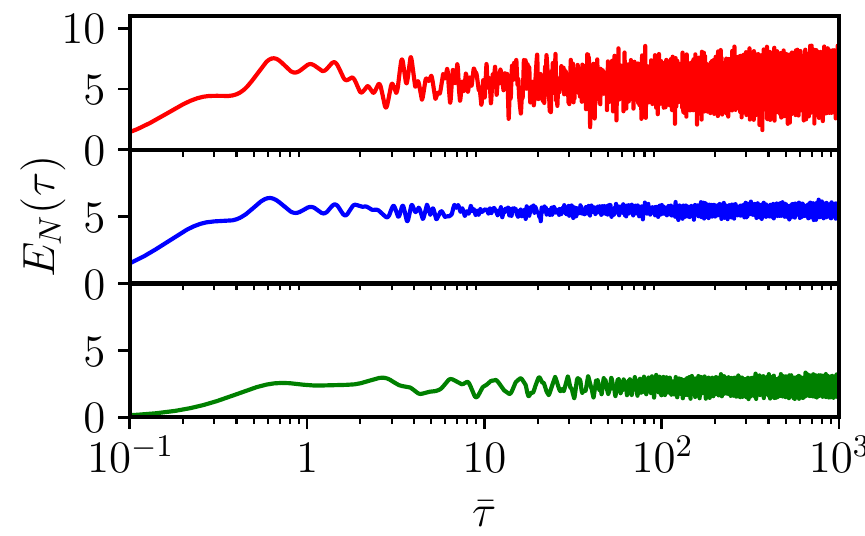}\put(4,80){}\end{overpic}
  \caption{(Color online) The mean local energy $E_{N}(\tau)$ as a function of the time $\tau$,
    for a disordered Heisenberg spin-chain QB operating in the various phases (from top to bottom,
    AL, MBL, ergodic). Here, and in the next two figures, the color code is the same used
    in Figs.~\ref{fig:Time} and~\ref{fig:ERG},
    while the Hamiltonian parameters for the representative cases corresponding to the three
    above mentioned phases are reported in this section.
    Data are for a single realization of disorder and $N=12$.}
  \label{fig:Time_XXZ}
\end{figure}

\begin{figure*}[t]
  \centering
  \vspace{1.em}
  \begin{overpic}[width=0.67\columnwidth]{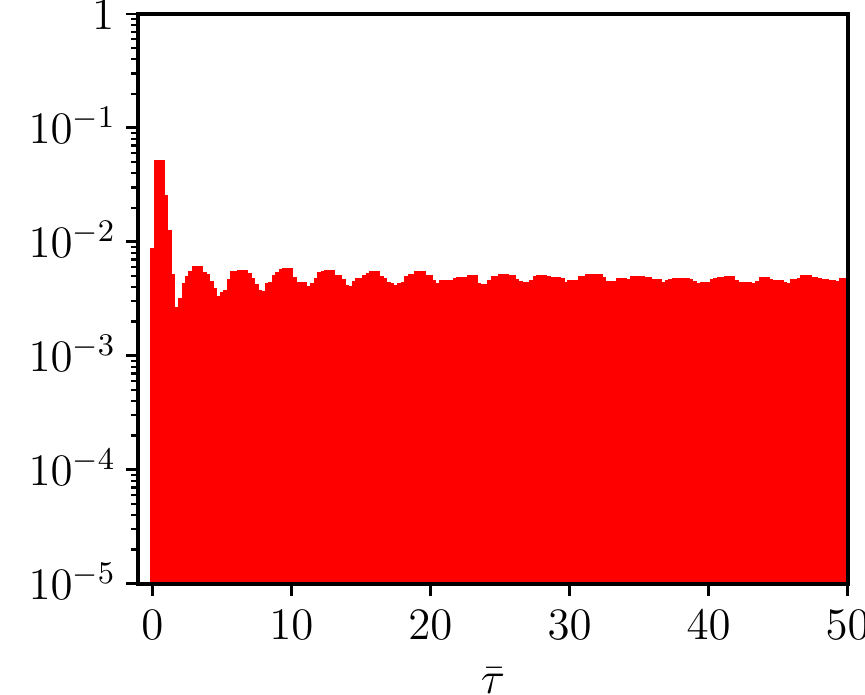}\put(2,84){\normalsize (a)}\end{overpic} 
  \begin{overpic}[width=0.67\columnwidth]{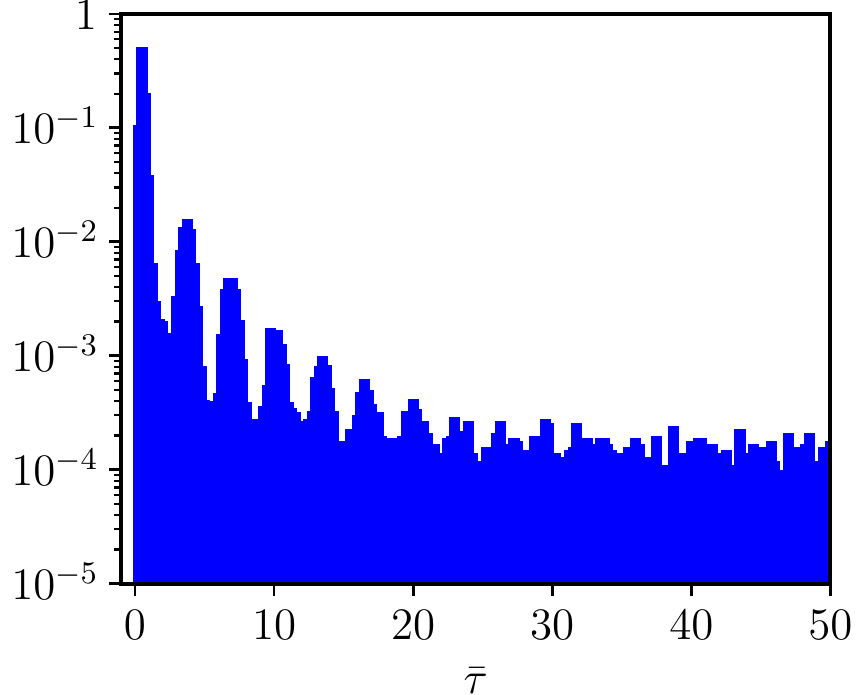}\put(2,84){\normalsize (b)}\end{overpic}
  \begin{overpic}[width=0.67\columnwidth]{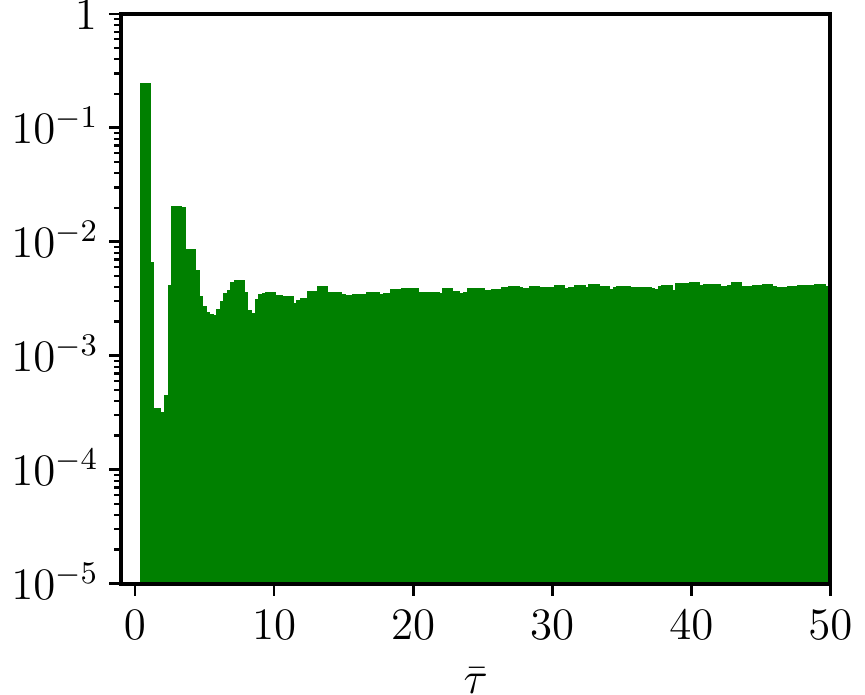}\put(2,84){\normalsize (c)}\end{overpic}\\
  \caption{(Color online) The probability $P(\bar{\tau})$ to find a certain value of the optimal charging time
    in a given phase, for a disordered Heisenberg spin-chain battery, is plotted as a function of $\bar{\tau}$.  
    Histograms in the three panels refer to a battery with $N=12$ units, in the above mentioned phases
    (from left to right, AL, MBL, ergodic), and have been obtained by averaging over $10^5$ realizations.
    The observation time is $\tau_{\rm max} = 50$.}
  \label{fig:Histo_XXZ}
\end{figure*}

The model defined above is usually referred to as the ``disordered XXZ
Heisenberg spin chain", and constitutes a prototypical
example where a many-body localization/delocalization transition
can be studied~\cite{Oganesyan07, Znidaric08, Pal10, Laflorencie15}.
Extensive numerical simulations have shown that, for $J = \Delta = 1$, such transition occurs
at $W^\star \approx 3.5$~\cite{Laflorencie15}.
We also recall that, by means of a Jordan-Wigner transformation
that maps spins into fermions, the model is unitarily equivalent to a one-dimensional
chain of spinless fermions, interacting through a nearest-neighbor
density-density term, and in the presence of a random onsite chemical potential.

The reason why we preferred to show results for the disordered quantum
Ising chain model in Sect.~\ref{sect:results} is that the latter only preserves a global $\mathbb{Z}_2$ symmetry
given by the parity operator $P = \prod_j \sigma^z_j$.
In contrast, the XXZ Heisenberg spin chain hosts a $U(1)$ symmetry associated with
the conservation of the total magnetization along the $\hat{\bm z}$ axis, i.e.,
\begin{equation}
  \big[ \mathcal{\tilde H}_0 + \mathcal{\tilde H}_1 , S^z \big] = 0~, \quad
  \mbox{where } \; S^z = \sum_j \sigma^z_j~,
\end{equation}
and therefore its dynamics is constrained onto a smaller manifold of the Hilbert space.
However, as we shall see below, the same qualitative features emerge.
As a consequence, we can reasonably assure that the MBL phase represents the optimal compromise,
in terms of stability properties and work extraction, where a many-body disordered QB can operate.

Analogously to the quantum Ising chain battery, by changing the various Hamiltonian parameters
of $\mathcal{\tilde H}_0$ and $\mathcal{\tilde H}_1$, one can probe the phase diagram of the model, which is characterized by 
the same three distinct phases analyzed before: the AL, MBL, and  ergodic phases.
We explicitly focus on one representative of each of such phases. However, we have checked that qualitatively analogous
features emerge when the values of the parameters are changed in such a way to stay within the same phase.
For the sake of simplicity, we express all the various parameters in units of $J=1$, which is thus
taken as the reference energy scale.
We set $\Delta = 0$ and $W=5$ to yield a representative point in the AL phase (red color in all the figures of this section);
$\Delta = 1$ and $W = 5$ for the MBL phase (blue color); $\Delta = 1$ and $W=1$ for the ergodic phase (green color).

The mean energy injected in the QB, for a single disorder realization, is shown in Fig.~\ref{fig:Time_XXZ}.
We observe that a QB operating in the AL phase is characterized by large temporal fluctuations of the energy,
which are generally suppressed when interactions are switched on
(not shown, for this model --- see, however, Fig.~\ref{fig:FLUCTUATION} for the quantum Ising chain).
While the energy at the optimal time $\bar{\tau}$ can be slightly larger in the AL phase, rather than
in the MBL phase, its time average is the same up to a $\lesssim 2 \%$ discrepancy.
In contrast, for the ergodic phase, a lower amount of disorder reflects into a smaller injected energy.

\begin{figure}[!t]
  \centering
  \vspace{1.em}
  \begin{overpic}[width=0.49\columnwidth]{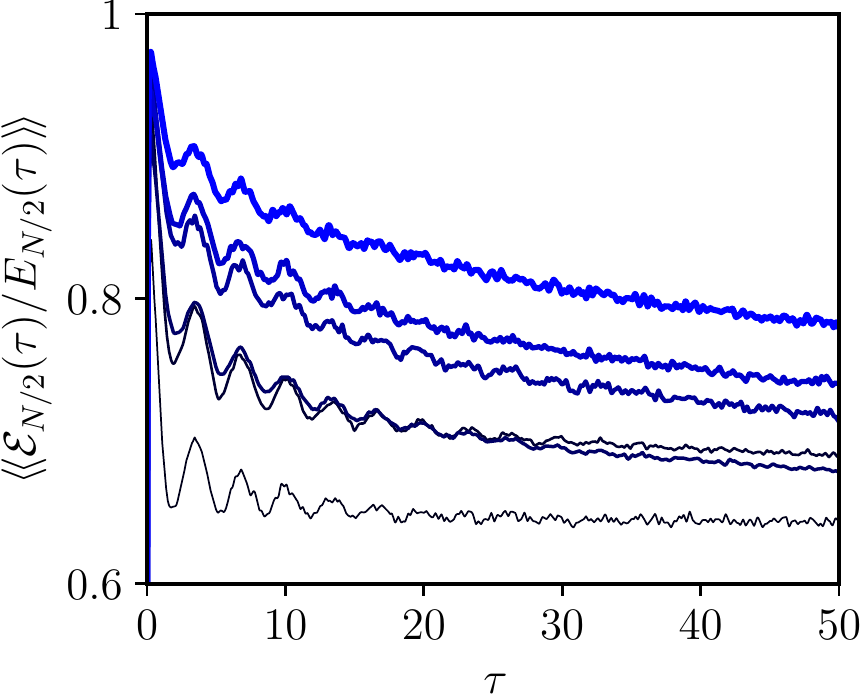}\put(2,84){\normalsize (a)}\end{overpic}
  \begin{overpic}[width=0.49\columnwidth]{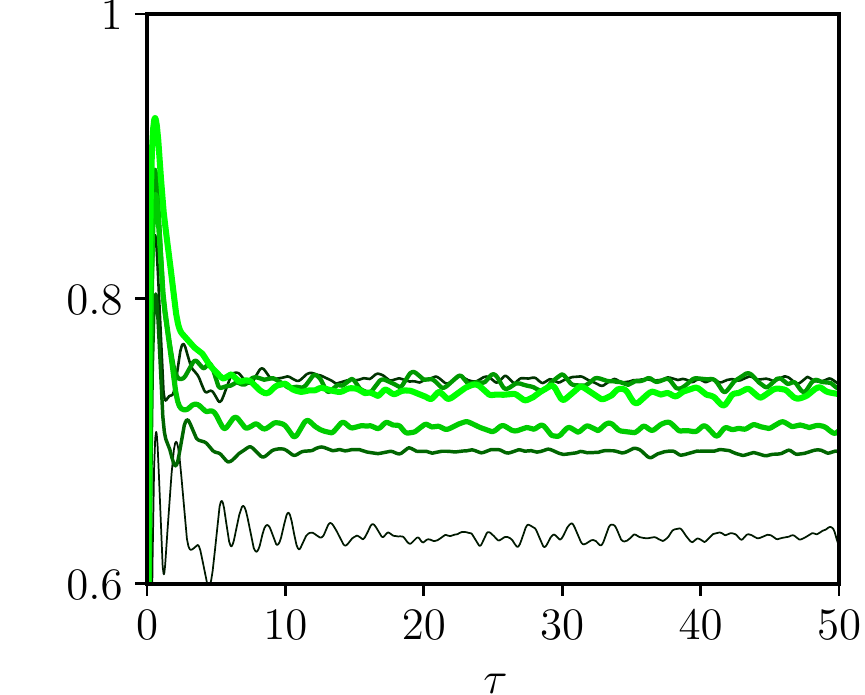}\put(2,84){\normalsize (b)}\end{overpic}
  \caption{(Color online) The fraction of extractable energy
    $\langle \! \langle \mathcal{E}_{N/2}(\tau)/{E}_{N/2}(\tau) \rangle \! \rangle$ as a function of $\tau$,
    for a disordered Heisenberg spin-chain battery. The two panels refer to MBL (a) and ergodic (b) phase.
    The various solid lines correspond to different values of the number $N$ of quantum cells: $N = 8,10,12,14,16,18$.
    The lines thicken and brighten with increasing $N$.
    Data are averaged over $10^3$ disorder realizations for $N$ up to $14$,
    and over $10^2$ realizations for $N=16$ and $18$.}
    \label{fig:ERG_XXZ}
\end{figure}

A closer look at the statistics of the optimal charging times $\bar{\tau}$ in the three quantum phases,
displayed in Fig.~\ref{fig:Histo_XXZ}, reveals the same qualitative features we observed in Sect.~\ref{subsec:times}.
Namely, after a pronounced peak for short times, both the AL phase [panel (a)]
and the ergodic phase [panel (c)] present an approximately flat distribution.
This witnesses the presence of temporal fluctuations at long times.
The MBL phase [panel (b)] is characterized by a rather distinct behavior:
the emerging sequence of peaks is reminiscent of those observed in the Ising chain model
[cf., Fig~\ref{fig:Time}, panel (d)].
The more spurious behavior observed here may be due to the reduced number $N$ of quantum
cells that we considered, and to the constraint in the quantum dynamics imposed by the
conservation of the magnetization along the $\hat{\bm z}$ axis.
Specifically, at $N=12$ as is the case for this figure, the quantum dynamics starting from the
ground state of $\mathcal{\tilde H}_0$ is constrained in a subspace of the system's Hilbert space
having dimension ${12 \choose 6} = 924$ (to be compared with the size of the Hilbert space $2^{12}=4096$).

The absence of complete ergodicity in the full Hilbert space, even in the ergodic phase,
also emerges in the analysis of the fraction of extractable energy, over half of the system.
We report the corresponding data for the MBL and the ergodic phase in Fig.~\ref{fig:ERG_XXZ}.
We clearly see, as observed in Sect.~\ref{subsec:extren}, that by increasing $N$, better performances in terms of work extraction
are achieved in a localized situation [panel (a)], rather than in the ergodic case [panel (b)]. 
However, we report a less distinct behavior for the two regimes, as compared with those observed
in the Ising model [cf., Fig.~\ref{fig:ERG}, panels (b) and (c)].


\begin{thebibliography}{77}
\bibitem{footnote1}
  By ``free" and ``local" we mean that the Hamiltonian $\mathcal{H}_0$ is a sum of terms which only involve a single quantum cell of the battery. For the sake of convenience and without loss of generality, we also set to zero the ground-state energy $\epsilon_{0}$ of $\mathcal{H}_0$, by adding to the Hamiltonian a trivial constant, i.e.~$+Nh$.
\bibitem{footnote2}
  For practical reasons, of relevance for both numerical calculations and real experiments, such maximum can be taken by calculating (measuring) $E_{N}(\tau)$ over a finite observation time window $\tau_{\rm max}$.
\bibitem{footnote3}
  We simulated systems with up to $N=18$ spins, over a maximum observation time window $\tau_{\rm max} = 100$. 
  We checked that, under these conditions, an integration time step $\delta t = 10^{-2}$
  is sufficient to guarantee convergence of all our results, with an error bar that is
  negligible on the scale of the plotted figures.
\end{thebibliography}
\end{document}